\documentclass[prd,nopacs,preprintnumbers,twocolumn,amsmath,nofootinbib,amssymb,floatfix]{revtex4}
\usepackage{graphicx,color,dcolumn,booktabs,bm}
\usepackage{longtable,lscape}
\usepackage{txfonts}
\usepackage{overpic}
\usepackage{amssymb}
\usepackage{epstopdf}
\usepackage{indentfirst}
\usepackage{feynmf}   
\usepackage{slashed}  
\usepackage{cases}
\usepackage{color}
\usepackage{float}
\usepackage{multirow}
\usepackage{epsfig,dsfont,amssymb,amsmath,amsfonts,amsbsy,mathrsfs}

\graphicspath{{Figures/}} %

\usepackage{hyperref}
\hypersetup{colorlinks,citecolor=blue,anchorcolor=red,menucolor=red, linkcolor=red,filecolor=red,runcolor=red,urlcolor=blue}

\makeatletter
\@addtoreset{equation}{section}
\makeatother

\allowdisplaybreaks

\begin{document}
\preprint{
	\vbox{
		\hbox{ADP-23-29/T1238}
}}

\title{ Structure of the $\mathbf{\Lambda(1670)}$ resonance}
\author{Jiong-Jiong Liu$^{1,2,3}$}
\author{Zhan-Wei Liu$^{1,2,3}$}\email{liuzhanwei@lzu.edu.cn}
\author{Kan Chen$^{4,5,6,7,8}$}
\author{Dan Guo$^{8}$}
\author{Derek B. Leinweber$^{9}$}\email{derek.leinweber@adelaide.edu.au}
\author{Xiang Liu$^{1,2,3}$}\email{xiangliu@lzu.edu.cn}
\author{Anthony W. Thomas$^{9}$}
\email{anthony.thomas@adelaide.edu.au}

\affiliation{
$^1$School of Physical Science and Technology, Lanzhou University, Lanzhou 730000, China\\
$^2$Research Center for Hadron and CSR Physics, Lanzhou University and Institute of Modern Physics of CAS, Lanzhou 730000, China\\
$^3$Lanzhou Center for Theoretical Physics, Key Laboratory of Theoretical Physics of Gansu Province, Key Laboratory of Quantum Theory and Applications of MoE, and MoE Frontiers Science Center for Rare Isotopes, Lanzhou University, Lanzhou 730000, China\\
$^4$School of Physics, Northwest University, Xi’an 710127, China\\
$^5$Shaanxi Key Laboratory for theoretical Physics Frontiers, Xi’an 710127, China\\
$^6$Institute of Modern Physics, Northwest University, Xi’an 710127, China\\
$^7$Peng Huanwu Center for Fundamental Theory, Xi’an 710127, China\\
$^8$School of Physics and State Key Laboratory of Nuclear Physics and Technology, Peking University, Beijing 100871, China\\
$^{9}$ARC Special Research Centre for the Subatomic Structure of Matter (CSSM), Department of Physics, University of Adelaide, Adelaide, South Australia 5005, Australia
}

\begin{abstract}
We examine the internal structure of the $\Lambda(1670)$ through an analysis of lattice QCD simulations and experimental data within Hamiltonian effective field theory. Two scenarios are presented. The first describes the $\Lambda(1670)$ as a bare three-quark basis state, which mixes with the $\pi\Sigma$, $\bar{K}N$, $\eta\Lambda$ and $K\Xi$ meson-baryon channels. In the second scenario, the $\Lambda(1670)$ is dynamically generated from these isospin-0 coupled channels. The $K^-p$ scattering data and the pole structures of the $\Lambda(1405)$ and the $\Lambda(1670)$ can be simultaneously described well in both scenarios. However, a comparison of the finite-volume spectra to lattice QCD calculations reveals significant differences between these scenarios, with a clear preference for the first case. Thus the lattice QCD results play a crucial role in allowing us to distinguish between these two scenarios for the internal structure of the $\Lambda(1670)$.
\end{abstract}

\maketitle

\section{introduction}\label{sec1}

Given that the strange quark is considerably heavier than the light quarks, it is a remarkable feature of the baryon spectrum that the lightest odd-parity baryon is not an excited state of the nucleon but lies within the $\Lambda$ family, with nonzero strangeness. 
This resonance, the $\Lambda(1405)$, with $I(J^P)=0(1/2^-)$, has been the subject of considerable theoretical work and speculation. Before the discovery of quarks, Dalitz and Tuan suggested that it might be a $\bar{K}N$ molecule, since its mass is slightly below the $\bar{K}N$ threshold~\cite{Dalitz:1960du}. However, there have been many other options explored since then, including conventional baryon states, dynamically generated states, three-quark states mixing with multiquark components as well as other explanations~\cite{Klempt:2009pi,Capstick:1986bm,Chen:2009de,Melde:2008yr,Bijker:2000gq,Glozman:1997ag,Loring:2001ky,Schat:2001xr,Menadue:2011pd,Engel:2012qp,An:2010wb,Hyodo:2011ur,Oller:2005ig,Borasoy:2005ie,Hyodo:2003qa,Jido:2003cb,GarciaRecio:2002td,Oller:2000fj,Oller:2006hx,Guo:2012vv,Zhang:2013cua,Zhang:2013sva,Geng:2007hz,Liu:2011sw,Xie:2013wfa,GarciaRecio:2003ks,Oset:2001cn,Xie:2016evi,Miyahara:2015cja,Pavao:2020zle,MartinezTorres:2012yi,Guo:2023wes}. 

In the first study based upon SU(3) chiral symmetry after the discovery of QCD, Veit {\it et al.} also concluded that their result ``strongly support[ed] the contention that the $\Lambda(1405)$ is a $\bar{K}N$ bound state''~\cite{Veit:1984jr}. 
This interpretation has also been supported by studies within the chiral unitary approach~\cite{Oller:2000fj,Jido:2003cb,Oller:2019opk}, as well as in an analysis of lattice QCD data by the CSSM group~\cite{Hall:2014uca,Hall:2016kou,Liu:2016wxq} using Hamiltonian effective field theory (HEFT)~\cite{Wu:2014vma,Hall:2013qba,Liu:2015ktc,Leinweber:2015kyz,Abell:2023nex,Yu:2023xxf}. 

There has been considerable interest in the analytic structure of the S-matrix for this system. In particular, the two-pole structure of the $\Lambda(1405)$, once SU(3) chiral symmetry of QCD was treated seriously, was unveiled for the first time in Ref. \cite{Oller:2000fj}. The two resonance poles in the second Riemann sheet are related to the thresholds of the $\bar{K}N$ and $\pi\Sigma$ channels~\cite{Mai:2020ltx,Hyodo:2020czb,Meissner:2020khl,Xie:2023cej,Liu:2016wxq,Lu:2022hwm}. The evolution of the poles for the $\Lambda(1380)$, $\Lambda(1405)$ and $\Lambda(1670)$ away from the SU(3) limit was first studied in 
Ref.~\cite{Jido:2003cb}.

There have recently been many other examples of systems which are suspected of being molecular in nature. For example, this has been proposed~\cite{Wang:2019krq,Chen:2019uvv} as an explanation of the  $\Xi(1620)$ recently observed  by the Belle collaboration~\cite{Sumihama:2018moz}. When it comes to heavy quarks, many multiquark states have been announced by different experiments, including the $P_c(4312)$, $P_c(4440)$, $P_c(4457)$~\cite{LHCb:2015yax,LHCb:2019kea}, $T_{cc}(3875)$~\cite{LHCb:2021auc,LHCb:2021vvq}, $P_{cs}(4459)$~\cite{LHCb:2020jpq}, and the 
$P_{cs}(4338)$~\cite{LHCb:2022ogu}. The occurrence of these exotic states near thresholds and their exotic quark content has resulted in the molecular picture being a popular interpretation of these states. 

As the $\Lambda(1405)$ is now commonly interpreted as a $\bar KN$ bound state, it is natural to ask where one might find the lightest $P$-wave $uds$ baryon expected in the conventional quark model. The analysis of Veit {\it et al.} suggested that the $J^P=1/2^-$ $\Lambda(1670)$ might be identified with this triquark core~\cite{Veit:1984jr}. In such a scenario the structure of the $\Lambda(1405)$ and the $\Lambda(1670)$ would be very different. It is important to investigate this interpretation very carefully, given the suggestion that once one removes molecularlike states from the baryon spectrum it appears as though the quark model idea of oscillatorlike major shells might be correct~\cite{Wu:2017qve}.

With this in mind, here we present a detailed study of the $S=-1,$ $J^P=\frac{1}{2}^-$ system, covering both the $\Lambda(1670)$ and the $\Lambda(1405)$ resonance region within HEFT. First we extend our earlier analysis of the cross section data for $K^-p$ scattering to include $K^-$ laboratory momenta up to 800 MeV/c, including the near threshold cross sections measured in 2022~\cite{Piscicchia:2022wmd}. The pole structure of the $\Lambda(1670)$ and the $\Lambda(1405)$ resonances are  examined. 

We analyze the lattice QCD data for the negative parity $\Lambda$ baryons~\cite{Engel:2012qp,Menadue:2011pd,Hall:2014uca,Gubler:2016viv}, as well as the corresponding eigenvectors describing the structure of those eigenstates. Given that just a few months ago, the BaSc collaboration presented their coupled-channel simulations with both single baryon and meson-baryon interpolating operators at $m_\pi\approx$200 MeV, which is close to the physical pion mass~\cite{BaryonScatteringBaSc:2023zvt,BaryonScatteringBaSc:2023ori}, we pay particular attention to their results~\cite{BaryonScatteringBaSc:2023zvt,BaryonScatteringBaSc:2023ori}. 

On the experimental side, there has been considerable progress in updating the $K^-p$ scattering information associated with the negative-parity $\Lambda$ baryons, which has been included in our analysis. For example, J-PARC provided the $\pi\Sigma$ invariant mass spectra in $K^-p$ induced reactions on the deuteron in 2022~\cite{J-PARCE31:2022plu}. The ALICE collaboration extracted the $K^-p$ scattering length with a measurement of momentum correlations in 2021~\cite{ALICE:2021szj}, while measurements of the energy shift of the 1s-state in kaonic-hydrogen by the SIDDHARTA collaboration have yielded precise values for the  $K^-p$ scattering lengths~\cite{Zmeskal:2015efj,SIDDHARTA:2011dsy}. Simultaneous $K^-p\to \Sigma^0\pi^0, \Lambda \pi^0$ near-threshold cross sections were measured at DA$\Phi$NE in 2022 \cite{Piscicchia:2022wmd}.

This paper is organized as follows. In Sec.~\ref{sec2}, we outline the HEFT framework as applied to the negative parity $\Lambda$ hyperons, while in Sec.~\ref{sec3} the corresponding numerical results and discussion are presented. A summary and concluding remarks are provided in Sec.~\ref{sec4}.

\section{Hamiltonian Effective Field Theory}
\label{sec2}
In this section, we introduce the framework within which we describe the $K^-p$ scattering processes. In order to obtain the mass spectra of $\Lambda$ baryons with $J^P=1/2^-$, which can be compared with those observed in lattice QCD, we also present the finite-volume Hamiltonian.

\subsection{The Hamiltonian}
In the rest frame, the Hamiltonian includes noninteracting and interacting parts
\begin{equation}
\label{Hamiltonian}
H^{I}=H^{I}_0+H^{I}_{int} \, ,
\end{equation}
where the kinetic energy piece of the Hamiltonian is written as
\begin{eqnarray}
\label{H0Hamiltonian}
H^{I}_0&=&\sum_{B_0}|B_0\rangle\, m_B^0\,\langle B_0| \nonumber\\
&&+\sum_{\alpha}\int d^3 k\left|\alpha\left(k\right)\right\rangle\left[\omega_{\alpha_{M}}\left(k\right)+\omega_{\alpha_B}\left(k\right)\right]\left\langle\alpha\left(k\right)\right|.\quad 
\end{eqnarray}
Here, $B_0$ denotes a ``bare baryon'' with mass $m_B^0$. This state may be interpreted as representing a quark model baryon, which is then dressed by its coupling to meson-baryon channels~\cite{Thomas:1982kv}. Here $\alpha$ labels the meson-baryon channel and $\alpha_M$ ($\alpha_B$) is the meson (baryon) in channel $\alpha$. The meson (baryon) energy is simply
\begin{eqnarray}
\label{freeenergy}
\omega_{\alpha_{M\left(B\right)}}\left(k\right)=\sqrt{m^2_{\alpha_{M\left(B\right)}}+k^2}.
\end{eqnarray}
For the interacting part of the Hamiltonian we include a vertex interaction coupling the bare baryon to the meson-baryon channel $\alpha$
\begin{eqnarray}
\label{bareHamiltonian}
g^{I}=\sum_{B_0,\alpha}\int d^3 k\left\{\left|B_0\right\rangle G^{I\dagger}_{B_0,\alpha}\left(k\right)\left\langle \alpha\left(k\right)\right|+h.c.\right\} \, ,
\end{eqnarray}
as well as the direct two-to-two particle interactions
\begin{eqnarray}
v^{I}=\sum_{\alpha,\beta}\int d^3 kd^3 k^\prime\left|\alpha(k)\right\rangle V^{I}_{\alpha,\beta}\left(k,k^\prime\right)\left\langle\beta(k^\prime)\right| \, .
\end{eqnarray}
The form factors associated with the coupling of the bare baryon to meson-baryon channel $\alpha$ are 
\begin{eqnarray}
\label{bareintaction}
G^{I}_{B_0,\alpha}(k)=\frac{\sqrt{3}\,g^I_{B_0,\alpha}}{2\pi f}\sqrt{\omega_{\alpha_M}(k)}\,u(k) \, ,
\end{eqnarray}
and the form of the potentials motivated by the Weinberg-Tomozawa  interaction~\cite{Weinberg:1966kf,Thomas:1981ps,Veit:1984jr} are
\begin{eqnarray}
V^{I}_{\alpha,\beta}\left(k,k^\prime\right)=g^I_{\alpha,\beta}\frac{\left[\omega_{\alpha_M}\left(k\right)+\omega_{\beta_B}\left(k^\prime\right)\right]u\left(k\right)u\left(k^\prime\right)}{8\pi^2f^2\sqrt{2\omega_{\alpha_M}\left(k\right)}
	\sqrt{2\omega_{\beta_M}\left(k^\prime\right)}} \, .
\label{interactionp}
\end{eqnarray}
Here, we use the dipole form factor $u(k)=(1+k^2/\Lambda^2)^{-2}$ with regulator parameter $\Lambda=1$ GeV. The $g_{B_0,\alpha}^I$ and $g^{I}_{\alpha,\beta}$ are the couplings of the corresponding interaction terms in the isospin $I$ channel.
As discussed in Sec.~III.~C of Ref.~\cite{Abell:2021awi}, when working in a
nonperturbative effective field theory, couplings encounter significant
renormalization such that the perturbative couplings are best promoted to fit
parameters. While this is a form of modeling, the L\"uscher formalism within HEFT
protects the model-independent relation between the scattering observables and the
finite-volume spectrum. As discussed in Sec.~\ref{sec:modelInDependence}, our task is to ensure there are sufficient parameters to accurately describe the scattering data.

\subsection{Infinite-volume scattering amplitude}\label{sec2b}
Here we introduce the formalism to describe the cross sections of $K^-p$ scattering in infinite volume. The two particle scattering T matrices can be obtained by the three-dimensional reduction of the coupled-channel Bethe-Salpeter equation
\begin{eqnarray}
T^{I}_{\alpha,\beta}(k,k^\prime;E)&=&\widetilde{V}^{I}_{\alpha,\beta}(k,k^\prime;E)\\
&&+\sum_\gamma
\int q^2dq\frac{\widetilde{V}^{I}_{\alpha,\gamma}(k,q;E)\, T^{I}_{\gamma,\beta}(q,k^\prime;E)}{E-\omega_{\gamma_1}(q)-\omega_{\gamma_2}(q)+i\epsilon},\nonumber
\end{eqnarray}
where the scattering potential can be expressed from the interaction Hamiltonian above
\begin{eqnarray}
\widetilde{V}^{I}_{\alpha,\beta}(k,k^\prime ;E)&=&\sum_{B_0}\frac{G^{I\dagger}_{B_0,\alpha}(k)\,G^{I}_{B_0,\beta}(k^\prime)}
{E-m_B^0}+V^I_{\alpha,\beta}(k,k^\prime) \, .
\end{eqnarray}
Note that for $K^-p$ scattering the T matrices, $t_{\alpha,\beta}(k,k^\prime;E)$, appear as a linear combination of $I=0$ and $I=1$ channels, i.e., $T_{\alpha,\beta}(k,k^\prime;E)=a\, T^{I=0}_{\alpha,\beta}(k,k^\prime;E)+b\, T^{I=1}_{\alpha,\beta}(k,k^\prime;E)$, where $a$ and $b$ involve the corresponding Clebsch-Gordan coefficients.
The poles associated with the $\Lambda(1405)$ and $\Lambda(1670)$ can be obtained by searching for those of the  $T^{I=0}_{\alpha,\beta}(k,k^\prime;E_{pole})$ matrix on the unphysical Riemann sheet.

In addition, the cross section for the process $\beta\rightarrow \alpha$ is related to the T matrices by
\begin{eqnarray}
\sigma_{\alpha,\beta}&=&\frac{4\pi^3 k_{\alpha\,{\rm cm}}\,\omega_{\alpha_M\,{\rm cm}}\,\omega_{\alpha_B\,{\rm cm}}\,\omega_{\beta_M\,{\rm cm}}\,\omega_{\beta_B\,{\rm cm}}}{E^2_{\rm cm}\,k_{\beta\,{\rm cm}}}\,\nonumber\\
&&\qquad \times |T_{\alpha,\beta}(k_{\alpha\,{\rm cm}},k_{\beta\,\rm cm};E_{\rm cm})|^2 \, , 
\label{crosssection}
\end{eqnarray}
where the subscript ``${\rm cm}$'' refers to the center-of-mass momentum frame.

\subsection{Finite-volume Hamiltonian model}
The scattering processes need interactions in both the $I=0$ and $I=1$ channels. Since the mass of $\Lambda(1670)$ is only 130 MeV below the $K\Xi$ threshold, in this work we include the  $K\Xi$ as well as the $\pi\Sigma$, $\bar{K}N$ and $\eta\Lambda$ channels for $I=0$ and the $\pi\Sigma$, $\bar{K}N$,  $\pi\Lambda$, $\eta\Sigma$, and $K\Xi$ channels for $I=1$.
In the finite volume, the Hamiltonian $\mathcal{H}$ consists of free and interacting terms, $\mathcal{H}=\mathcal{H}_0+\mathcal{H}_I$. Such a Hamiltonian can be expressed as a matrix, using the corresponding discrete momentum basis.

In the cubic, finite volume of lattice QCD with length $L$, the momentum of a particle is $k_n=2\pi\sqrt{n}/L$, with $n=n_x^2+n_y^2+n_z^2$, where $n=0,1,2,...$.
The noninteracting isospin-0 Hamiltonian is
%
\begin{eqnarray}
\mathcal{H}_{0}^{0}&=&\text{diag} \left\{ m_B^0,\omega_{\pi\Sigma}\left(k_0\right),\omega_{\bar{K}N}
\left(k_{0}\right),\omega_{\eta\Lambda}\left(k_0\right), \right. \nonumber \\
& & \left. \omega_{K\Xi}\left(k_0\right),...,
\omega_{\pi\Sigma}\left(k_1\right),... \right\}
\end{eqnarray}
%
and the interacting Hamiltonian can be written as
\begin{widetext}
\begin{eqnarray}
	\mathcal{H}^0_{int}=\left(
	\begin{array}{cccccccccc}
		0 & \mathcal{G}^0_{B_0,\pi\Sigma}\left(k_0\right) & \mathcal{G}^0_{B_0,\bar{K}N}\left(k_0\right) & \mathcal{G}^0_{B_0,\eta\Lambda}\left(k_0\right) & \mathcal{G}^0_{B_{0},K\Xi}\left(k_0\right) & \mathcal{G}^0_{B_0,\pi\Sigma}\left(k_1\right) & \cdots \\
		\mathcal{G}^0_{B_0,\pi\Sigma}\left(k_0\right) & \mathcal{V}^0_{\pi\Sigma,\pi\Sigma}\left(k_0,k_0\right) & \mathcal{V}^0_{\pi\Sigma,\bar{K}N}\left(k_0,k_0\right) & \mathcal{V}^0_{\pi\Sigma,\eta\Lambda}\left(k_0,k_0\right) & \mathcal{V}^0_{\pi\Sigma,K\Xi}\left(k_0,k_0\right) &  \mathcal{V}^0_{\pi\Sigma,\pi\Sigma}\left(k_0,k_1\right) & \cdots \\
		\mathcal{G}^0_{B_0,\bar{K}N}\left(k_0\right) & \mathcal{V}^0_{\bar{K}N,\pi\Sigma}\left(k_0,k_0\right) & \mathcal{V}^0_{\bar{K}N,\bar{K}N}\left(k_0,k_0\right)&\mathcal{V}^0_{\bar{K}N,\eta\Lambda}\left(k_0,k_0\right) & \mathcal{V}^0_{\bar{K}N,K\Xi}\left(k_0,k_0\right) & \mathcal{V}^0_{\pi\Sigma,\pi\Sigma}\left(k_0,k_1\right) & \cdots  \\
		\mathcal{G}^0_{B_0,\eta\Lambda}\left(k_0\right) & \mathcal{V}^0_{\eta\Lambda,\pi\Sigma}\left(k_0,k_0\right) & \mathcal{V}^0_{\eta\Lambda,\bar{K}N}\left(k_0,k_0\right) & \mathcal{V}^0_{\eta\Lambda,\eta\Lambda}\left(k_0,k_0\right) & \mathcal{V}^0_{\eta\Lambda,K\Xi}\left(k_0,k_0\right) &  \mathcal{V}^0_{\eta\Lambda,\pi\Sigma}\left(k_0,k_1\right) & \cdots \\
		\mathcal{G}^0_{B_0,K\Xi}\left(k_0\right) & \mathcal{V}^0_{K\Xi,\pi\Sigma}\left(k_0,k_0\right) & \mathcal{V}^0_{K\Xi,\bar{K}N}\left(k_0,k_0\right) & \mathcal{V}^0_{K\Xi,\eta\Lambda}\left(k_0,k_0\right) & \mathcal{V}^0_{K\Xi,K\Xi}\left(k_0,k_0\right)&  \mathcal{V}^0_{K\Xi,\pi\Sigma}\left(k_0,k_1\right) & \cdots \\
		\mathcal{G}^0_{B_0,\pi\Sigma}\left(k_1\right) & \mathcal{V}^0_{\pi\Sigma,\pi\Sigma}\left(k_1,k_0\right) & \mathcal{V}^0_{\pi\Sigma,\bar{K}N}\left(k_1,k_0\right) & \mathcal{V}^0_{\pi\Sigma,\eta\Lambda}\left(k_1,k_0\right) & \mathcal{V}^0_{\pi\Sigma,K\Xi}\left(k_1,k_0\right) & \mathcal{V}^0_{\pi\Sigma,\pi\Sigma}\left(k_1,k_1\right) & \cdots \\
		\vdots & \vdots & \vdots & \vdots & \vdots & \vdots & \ddots  \\
	\end{array}
	\right),
\end{eqnarray}
\end{widetext}
where
\begin{eqnarray}
\mathcal{G}^0_{B_0,\alpha}&=&\sqrt{\frac{C_3\left(n\right)}{4\pi}}\left(\frac{2\pi}{L}\right)^{3/2}G_{B_0,\alpha}^0\left(k_n\right),\\
\mathcal{V}^0_{\alpha,\beta}\left(k_n,k_m\right)&=&\frac{\sqrt{C_3\left(n\right)C_{3}\left(m\right)}}{4\pi}\left(\frac{2\pi}{L}\right)^3V^0_{\alpha,\beta}\left(k_n,k_m\right) \, .\qquad
\end{eqnarray}
Here, $C_{3}(n)$ denotes the number of ways one can sum the squares of three integers to equal $n$.

\section{ Model (in)dependence in HEFT}
\label{sec:modelInDependence}

Understanding the model-dependent and model-independent aspects of
HEFT is important. HEFT incorporates the L\"uscher formalism
\cite{Wu:2014vma,Hall:2013qba}, and therefore there are aspects of the
calculation that share the same level of model independence as the
L\"uscher formalism itself.

\subsection{Model independence}

The L\"uscher formalism provides a rigorous relationship between the
finite-volume energy spectrum and the scattering amplitudes of
infinite-volume experiment.  In HEFT, this relationship is mediated by
a Hamiltonian.  In the traditional approach, the parameters of the
Hamiltonian are tuned to describe lattice QCD results.  When the fit
provides a high-quality description of lattice QCD results, the
associated scattering-amplitude predictions are of high quality.  The
key is to have a sufficient number of tunable parameters within the
Hamiltonian to accurately describe the lattice QCD results.

However, in the baryon sector, high-quality lattice QCD results are
scarce and HEFT is usually fit to experimental data first. The HEFT
formalism then describes the finite-volume dependence of the baryon
spectrum, indicating where high-precision lattice QCD results will
reside.  This is the approach adopted herein.  We will show
high-quality fits to the experimental scattering observables such that
HEFT provides rigorous predictions of the finite-volume lattice QCD
spectrum with model independence at the level of the L\"uscher
formalism.

Of course, this model independence is restricted to the case of
matched quark masses in finite-volume and infinite-volume.  The
L\"uscher formalism provides no avenue for changing the quark mass.
In other words, direct contact with lattice QCD results is only
possible when the quark masses used in the lattice QCD simulations are
physical.

On the other hand, $\chi$PT is renowned for describing the quark mass
dependence of hadron properties in a model-independent manner,
provided one employs the truncated expansion in the power-counting
regime, where higher-order terms not considered in the expansion are
small by definition.  Given that finite-volume HEFT reproduces the leading behavior of
finite-volume $\chi$PT in the perturbative limit by construction
\cite{Hall:2013qba,Abell:2021awi}, it is reasonable to explore the
extent to which this model independence persists in the full
nonperturbative calculation of HEFT.

This has been explored in Ref.~\cite{Abell:2021awi}.  In the one-channel
case where a single particle basis state (e.g. a quark-model like
$\Delta$) couples to one two-particle channel (e.g. $\pi N$), the
independence of the results on the form of regularisation is
reminiscent of that realised in $\chi$PT.  Any change in the regulator is
absorbed by the low-energy coefficients such that the renormalised
coefficients are physical, independent of the renormalisation scheme. 

However, in the more complicated two-channel case with a $\pi \Delta$
channel added, the same was not observed.  The form of the Hamiltonian
becomes constrained, describing experimental data accurately for only
a limited range of parameters with specific regulator shapes.  The
Hamiltonian becomes a model in this case, with regulator-function
scales and shapes governed by the experimental data.  The principles
of chiral {\em perturbation} theory no longer apply in this
nonperturbative calculation.  However, for fit parameters that
describe the data well, the model independence of the L\"uscher
formalism remains intact.  The Hamiltonian is only a mediary.

\subsection{Quark mass variation}

The consideration of variation of the quark masses away from the
physical point provides further constraints on the Hamiltonian.  In
particular, lattice QCD results away from the physical point provide
new constraints on the form of the Hamiltonian.  In the two-channel
case, the Hamiltonian becomes tightly constrained when considering
experimental scattering data and lattice QCD results together
\cite{Abell:2021awi}.

With the Hamiltonian determined by one set of lattice results, one can
then make predictions of the finite-volume spectrum considered by
other lattice groups at different volumes and different quark masses.
This is a central aim of the current investigation where we confront very recent lattice QCD predictions for the odd-parity
$\Lambda$ spectrum at an unphysical pion mass of 204 MeV~\cite{BaryonScatteringBaSc:2023zvt,BaryonScatteringBaSc:2023ori}.  The
Hamiltonian will be constrained by lattice QCD results from
Refs.~\cite{Menadue:2011pd,Hall:2014uca} for the lowest-lying
odd-parity excitation, such that the confrontation with contemporary
lattice QCD results is predictive.

For the cases previously considered in the baryon spectrum the
predictions of HEFT are in agreement with lattice QCD spectrum
predictions.  For example, in the $\Delta$-channel, HEFT successfully
predicts the finite-volume spectrum of the CLS consortium
\cite{Abell:2021awi,Morningstar:2021ewk}.  In the $N1/2^+$ channel,
HEFT reproduces the lattice QCD results from Lang {\it et al.}
\cite{Wu:2017qve,Lang:2016hnn}.  In the $N1/2^-$ channel, HEFT
successfully predicts spectra from the CLS consortium
\cite{Abell:2023nex,Bulava:2022vpq}, the HSC
\cite{Abell:2023nex,Edwards:2011jj,Edwards:2012fx} and Lang \&
Verducci \cite{Abell:2023nex,Lang:2012db}.  Thus one concludes that
the systematic errors of the HEFT approach to quark-mass variation are
small on the scale of contemporary lattice QCD uncertainties.  As the
Hamiltonian is constrained by model-independent scattering data and
lattice QCD results, we expect this success to be realised in the
current investigation.

Variation in the quark mass is conducted in the same spirit as for
$\chi$PT.  The couplings are held constant and the hadron masses
participating in the theory take values as determined in lattice QCD.
The single-particle bare basis state acquires a quark mass dependence
and this is done in the usual fashion by drawing on terms analytic in
the quark mass.  In most cases, lattice QCD results are only able to
constrain a term linear in $m_\pi^2$, as is the case here.

The model independence associated with the movement of quark masses
away from the physical point is largely governed by the distance one
chooses to move from the physical quark-mass point. The HEFT approach
is systematically improvable, reliant on high-quality lattice QCD
results to constrain the higher-order terms that one can introduce.
For example, one could include an additional analytic $m_\pi^4$ term
or higher-order interaction terms from the chiral Lagrangian.
However, this increased level of precision is not yet demanded by
current experimental measurements nor contemporary lattice QCD
results.

\subsection{Model dependence}

Now that the Hamiltonian has become a tightly constrained
model, the eigenvectors describing the manner in which the
noninteracting basis states come together to compose the eigenstates
of the spectrum are model dependent. At the same time, there is little
freedom in the model parameters of the Hamiltonian such that the
predictions of the Hamiltonian are well defined.

The information contained in the Hamiltonian eigenvectors describing
the basis-state composition of finite-volume energy eigenstates is
analogous to the information contained within the eigenvectors of
lattice QCD correlation matrices describing the linear combination of
interpolating fields isolating energy eigenstates on the lattice.
These too are model dependent, governed by the nature of the
interpolating fields used to construct the correlation matrix.  

What is remarkable is that with a suitable renormalisation scheme on
the lattice (e.g. interpolators are normalised to set diagonal
correlators equal to 1 at one time slice after the source), the
composition of the states drawn from the lattice correlation matrix is
very similar to the description provided by HEFT
\cite{Wu:2017qve,Abell:2023nex}. While both eigenvector sets are model
dependent, their similarity does indeed provide some relevant insight
into hadron structure.  And because regularisation in the Hamiltonian
is tightly constrained, one can begin to separate out the
contributions of bare versus two-particle channels.

\subsection{Error analysis}

It may be of interest to compare the systematic uncertainties of the
HEFT formalism with the statistical uncertainties of contemporary
lattice QCD determinations of the finite-volume hadron spectrum.  To
do so requires an exploration of alternative Hamiltonians that
continue to describe both experimental data and lattice QCD results.

Variation of the regularisation parameters provides an opportunity to
move in the Hamiltonian parameter space.  However, the constraints of
experiment and lattice QCD are quite effective in constraining the
Hamiltonian, allowing only a small range of variation in the
regularisation parameters. Moving the parameters outside of the range
allowed by the data, spoils the fit to the data and thus the
associated finite-volume energy spectrum.  

Recalling that the embedded L\"uscher formalism governs the relation
between the scattering data and the finite-volume spectrum and noting
that the Hamiltonian plays only a mediary role, one concludes that
the systematic error is governed by the quality of the experimental
data and its ability to uniquely constrain the multichannel
Hamiltonian. This issue is quantified in the following section.

\subsection{Summary}

In summary, there is a direct model-independent link between the
scattering observables of experiment and the finite-volume spectrum
calculated in HEFT at physical quark masses.  This model independence
is founded on the L\"uscher formalism embedded with HEFT. Similarly,
variation of the quark masses away from the physical quark mass has
systematic uncertainties that are small relative to contemporary
lattice QCD spectral uncertainties.  Finally, the Hamiltonian
eigenvectors describing the basis-state composition of finite-volume
energy eigenstates are model dependent.  They are analogous to the
interpolator dependent eigenvectors of lattice QCD correlation
matrices describing the linear combination of interpolating fields
isolating energy eigenstates on the lattice.  The similarity displayed
by these two different sets of eigenvectors suggests that they do
indeed provide insight into hadron structure.


\section{Numerical Results}\label{sec3}
In this section, we first study the $K^-p$ cross section, adjusting the parameters of the interaction Hamiltonian to reproduce the experimental data. Then we use these fitted parameters to discuss the finite-volume results.

\subsection{Cross section}
As the internal structure of the $\Lambda(1670)$ is still not very clear, we examine two possible scenarios. In the first, we postulate that the $\Lambda(1670)$ is a resonance which is dynamically generated through rescattering in the isospin-0 $\pi\Sigma$, $\bar{K}N$, $\eta\Lambda$ and $K\Xi$ coupled channels. In the second, the $\Lambda(1670)$ is treated as a quark-model-like baryon, which mixes with the meson-baryon channels. We fit the experimental data for the $K^-p\rightarrow K^-p$, $K^-p\rightarrow \bar{K}^0n$, $K^-p\rightarrow \pi^0\Lambda^0$, $K^-p\rightarrow \pi^-\Sigma^+$, $K^-p\rightarrow \pi^0\Sigma^0$, $K^-p\rightarrow \pi^+\Sigma^-$, and $K^-p\rightarrow \eta\Lambda$ cross sections over the laboratory momentum range 0-800 MeV, which covers both resonances. We will explore which scenario can give a better description to the experimental cross section data.
\begin{figure}[htbp]
\flushleft
\includegraphics[width=4.3cm]{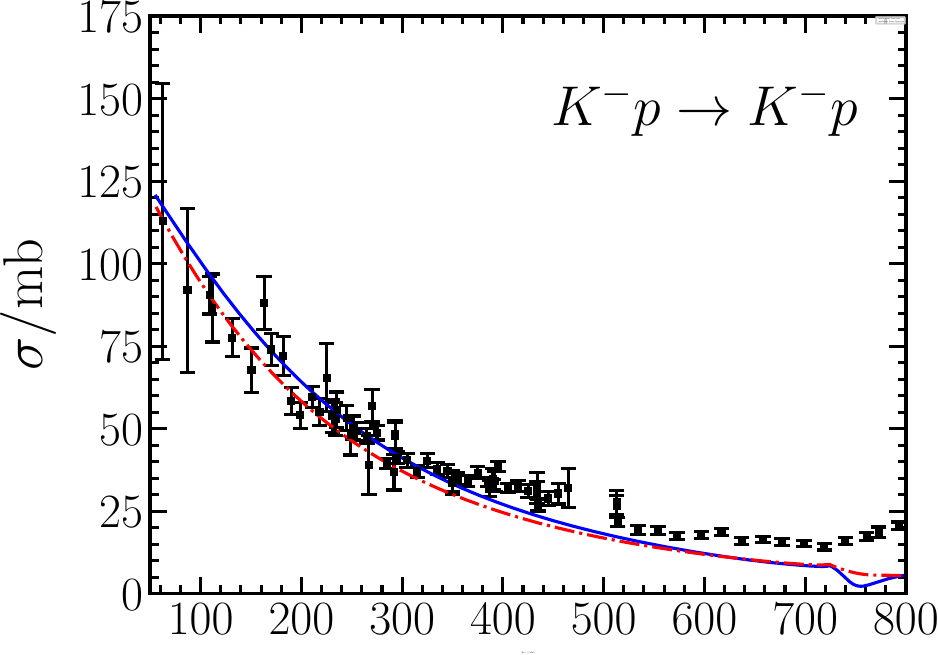}
\includegraphics[width=4.0cm]{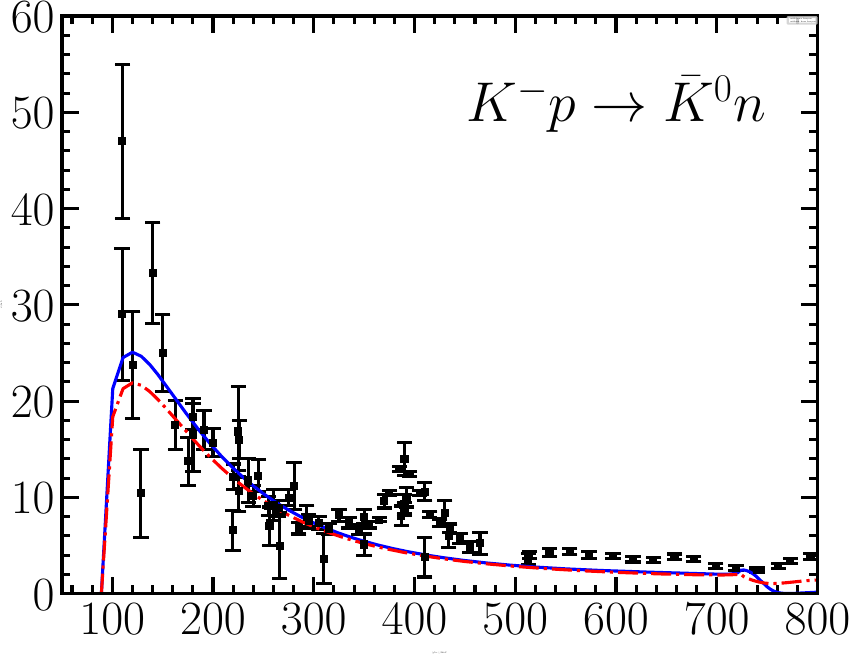}
\includegraphics[width=4.3cm]{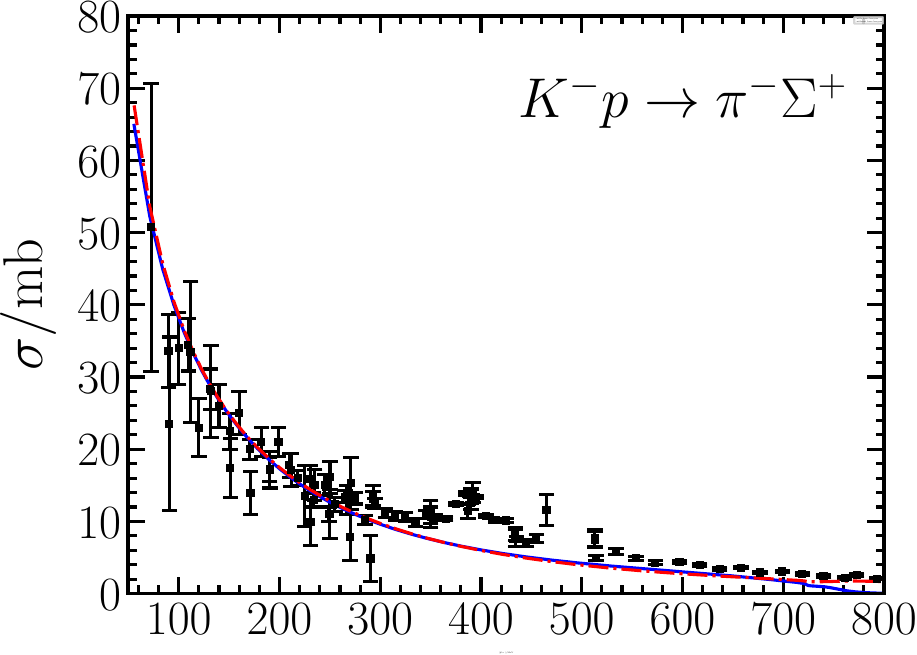}
\includegraphics[width=4.05cm]{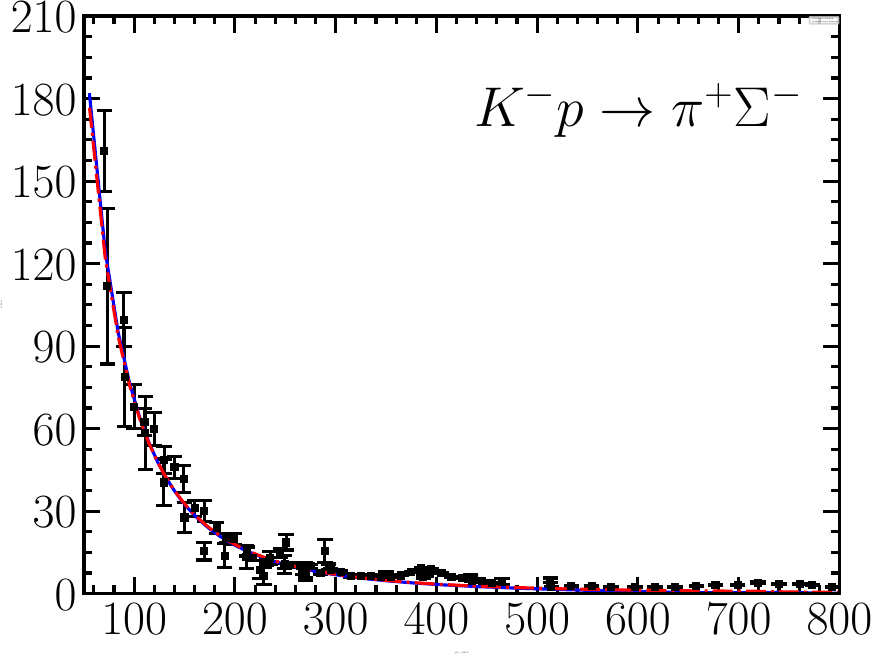}
\includegraphics[width=4.3cm]{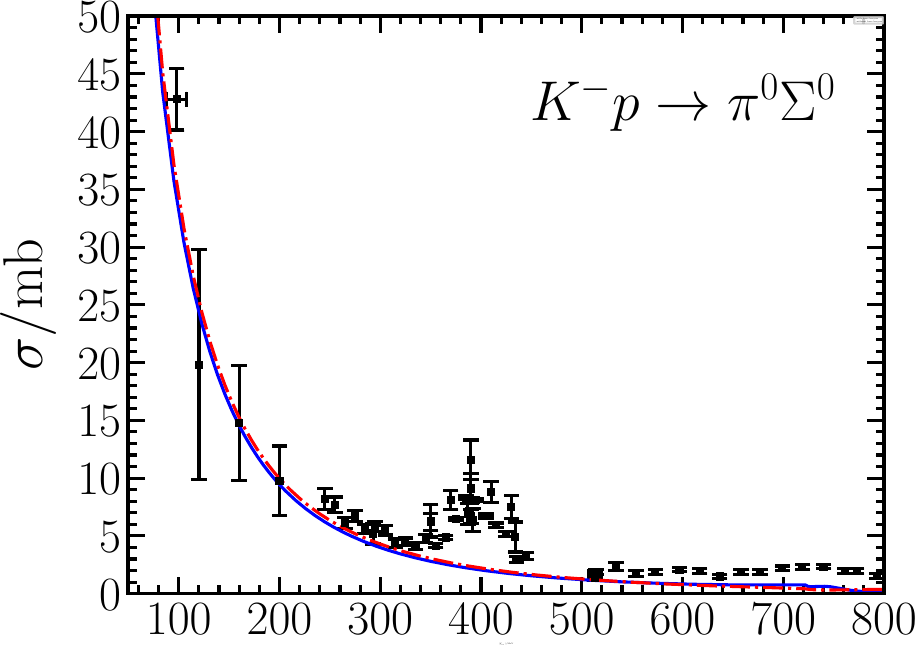}
\includegraphics[width=4.0cm]{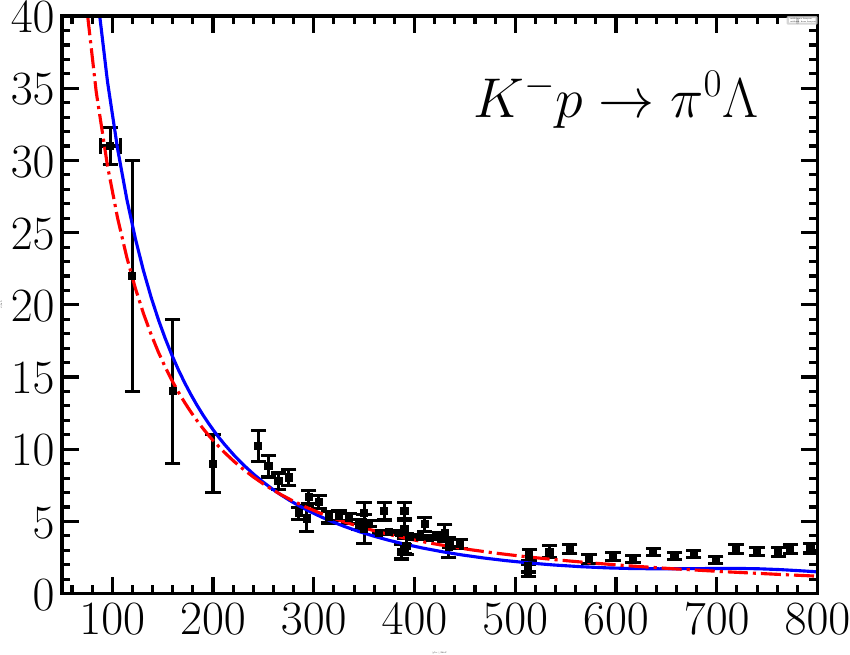}
\includegraphics[width=4.3cm]{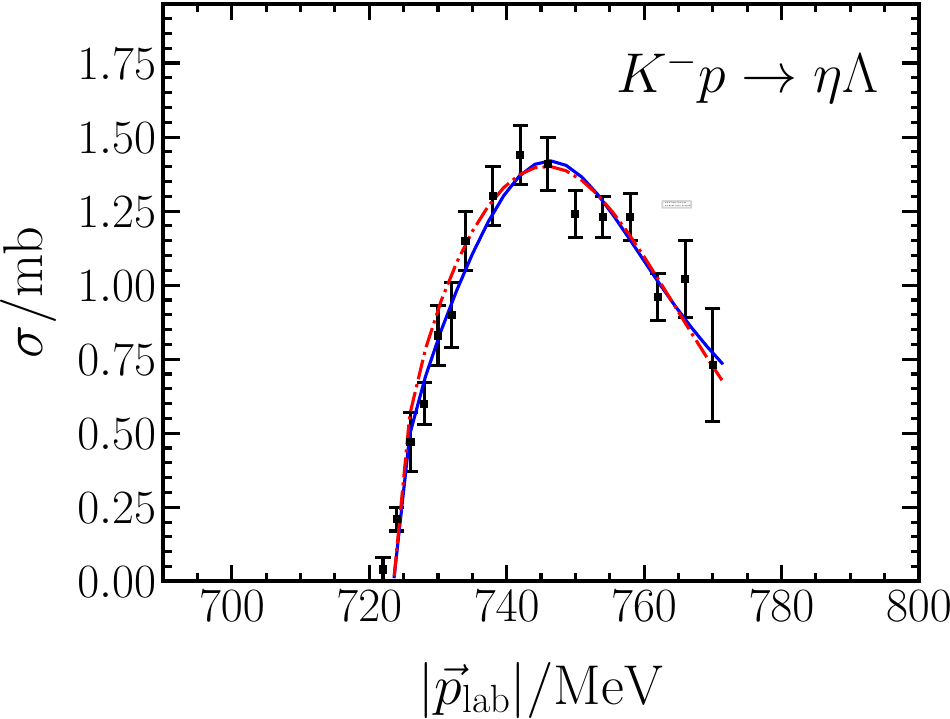}
\includegraphics[width=4.2cm]{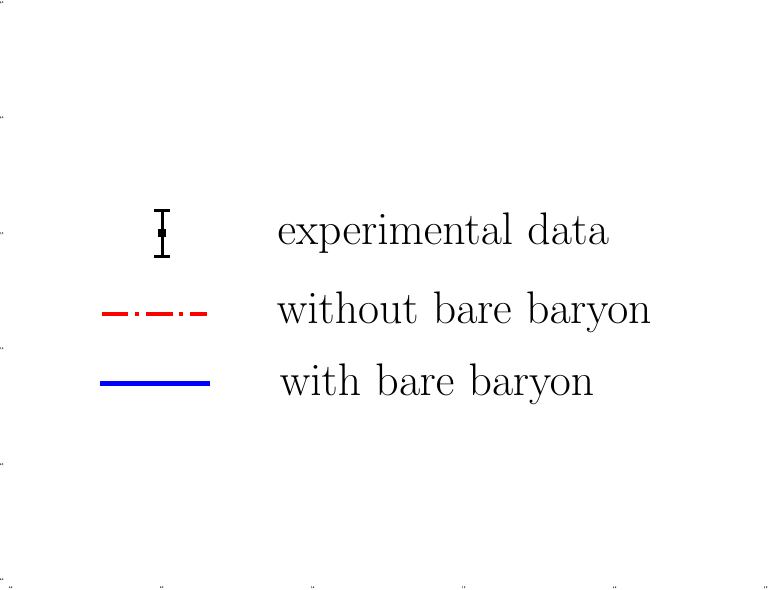}

\caption{Experimental data \cite{Abrams:1965zz,Sakitt:1965kh,Kim:1965zzd,Mast:1975pv,Bangerter:1980px,Ciborowski:1982et,Evans:1983hz,Mast:1974sx,Nordin:1961zz,Berley:1996zh,
		FerroLuzzi:1962zza,Watson:1963zz,Eberhard:1959zz,Piscicchia:2022wmd} for the cross sections of $K^-p\rightarrow K^-p$, $\bar{K}^0n$, $\pi^0\Lambda^0$, $\pi^-\Sigma^+$, $\pi^0\Sigma^0$, $\pi^+\Sigma^-$, and $\eta \Lambda$ scattering processes. The blue solid lines denote the fit including a bare quark-model-like basis state, while the red dashed lines denote the fit without including a bare-baryon component. The peaks around 400 MeV are associated with the $D$-wave $\Lambda(1520)$ state which is not considered in this work as described in the text.}\label{fitting}
		\end{figure}

		We present a comparison of the experimental cross sections with our fitted results in Fig.~\ref{fitting}. The fitted parameters are given in Table~\ref{coupling}. The peak of $\Lambda(1670)$ can be clearly seen in the $K^-p\rightarrow \eta\Lambda$ channel shown in the last subfigure of Fig.~\ref{fitting}. It is clear that we can describe this resonance well, both with and without the bare quark-model-like baryon. 
		
		From Fig.~\ref{fitting}, we see that the calculated cross section and the experimental cross section data have clear discrepancies at laboratory momenta  $\vec{p}_{\textrm{lab}}=350\sim450$ MeV in the $K^-p\rightarrow$ $\bar{K}^0n$, $\pi^-\Sigma^+$, and $\pi^0\Sigma^0$ processes. Here, we need to emphasize that we only introduce the $S$-wave interactions relevant to spin-1/2 odd-parity $\Lambda$ baryons. It is well known (e.g., Refs.~\cite{Lutz:2001yb,He:1993et}) that such discrepancies can be well understood by introducing the effects from $P$ and $D$ waves, while the $S$-wave interactions play essentially no role in forming those peaks. Specifically, 
		Refs.~\cite{He:1993et,Zhong:2013oqa,Zhong:2008km} presented detailed analyses showing that the peaks at $p_{\textrm{lab}}=350\sim450$ MeV in $K^-p\rightarrow \pi^0\Sigma^0$ and $\pi^{\pm}\Sigma^{\mp}$ arise mainly from contributions involving the $D$-wave $\Lambda(1520)$ state. The contribution from the $\Lambda(1520)$ is also responsible for the peak in the energy range $\vec{p}_{\textrm{lab}}=350\sim450$ MeV in the $K^-p\rightarrow \bar{K}^0n$ process \cite{Zhong:2013oqa}. Because they have different quantum numbers, these contributions cannot influence the $\Lambda(1670)$, and we do not consider them further.
		
		In the channels of $K^-p\rightarrow K^-p$, $K^-p\rightarrow \bar{K}^0n$, $K^-p\rightarrow \pi^0\Lambda^0$, and $K^-p\rightarrow \pi^0\Sigma^0$, the cross sections for $\vec{p}_{\textrm{lab}}>500$ MeV are much smaller than  those at LOW momenta. The $\Lambda(1670)$ resonance is not obvious in these channels either. Since we do not consider further channels, such as $\bar K^* N$ and $\pi \Sigma^*$,  which are close to this energy region, our results deviate from the experimental data for $\vec{p}_{\textrm{lab}}>500$ MeV in some channels.
		
		Except for these minor discrepancies, our calculations in both scenarios give a very good description of the experimental data. The recently measured threshold cross sections in Ref.~\cite{Piscicchia:2022wmd} for $K^-p\rightarrow \pi^0\Lambda^0$, and $K^-p\rightarrow \pi^0\Sigma^0$ provide more accurate constraints and can be described well, as can be seen in the third row of 
		Fig.~\ref{fitting}. The line shapes of the fits with and without the inclusion of a bare-baryon contribution are very similar. 
		\renewcommand\tabcolsep{0.3cm}
		\renewcommand{\arraystretch}{1.8}
		\begin{table}[!tbp]
\caption{The fit parameters obtained from $K^-p$ cross sections within the following two scenarios. One describes the $\Lambda(1670)$ as a bare quark-model-like single-particle state mixed with meson-baryon interactions from the $\pi\Sigma$, $\bar{K}N$, $\eta\Lambda$, and $K\Xi$ channels. The other describes the $\Lambda(1670)$ as pure dynamically-generated resonance from isoscalar coupled channels.
	Error estimates for the bare baryon case are obtained through the consideration of allowed variation in the regularisation parameter, $\Lambda$, as described in Sec~\ref{sec:uncertaintyAnalysis}. 
}
\centering
\label{coupling}
\begin{tabular}{cccccccccccc}
	\toprule[0.3pt]
	\hline
	\hline
	Coupling& Without bare baryon&With bare baryon\\
	\hline
	$\Lambda$ (GeV)&1.0&$1.0^{+0.1}_{-0.1}$\\
	$g^0_{\bar{K}N,\bar{K}N}$&-2.108&$-2.180^{+0.280}_{-0.135}$\\
	$g^0_{\bar{K}N,\pi\Sigma}$&0.837&$0.620^{-0.076}_{+0.080}$\\
	$g^0_{\bar{K}N,\eta\Lambda}$&-0.461&$-0.472^{+0.471}_{-0.864}$\\
	$g^0_{\pi\Sigma,\pi\Sigma}$&-1.728&$-1.200^{-0.078}_{-0.116}$\\
	$g^0_{\pi\Sigma,K\Xi}$&-0.001&$-1.800^{+0.452}_{-0.200}$\\
	$g^0_{\eta\Lambda,K\Xi}$&0.835&$1.993^{-0.668}_{-0.047}$\\
	$g^0_{K\Xi,K\Xi}$&-3.393&$-1.000^{-0.001}_{-2.314}$\\
	
	$g^1_{\bar{K}N,\bar{K}N}$&-0.028&$-0.001^{+0.000}_{+0.000}$\\
	$g^1_{\bar{K}N,\pi\Sigma}$&0.829&$0.985^{-0.152}_{+0.268}$\\
	$g^1_{\bar{K}N,\pi\Lambda}$&0.001&$0.990^{-0.108}_{+0.154}$\\
	$g^1_{\bar{K}N,\eta\Sigma}$&1.557&$1.500^{-0.212}_{+0.215}$\\
	$g^1_{\pi\Sigma,\pi\Sigma}$&-1.351&$-0.001^{-0.001}_{+0.000}$\\
	$g^1_{\pi\Sigma,K\Xi}$&-1.017&$-1.341^{+0.197}_{-0.439}$\\
	$g^1_{\pi\Lambda,K\Xi}$&2.904&$0.011^{-0.010}_{-0.010}$\\
	$g^1_{\eta\Sigma,K\Xi}$&4.690&$0.001^{+0.000}_{+0.000}$\\
	$g^1_{K\Xi,K\Xi}$&-0.447&$-3.700^{+0.660}_{-0.840}$\\
	
	$g^0_{B_0,\bar{K}N}$&-&$0.091^{-0.014}_{+0.032}$\\
	$g^0_{B_0,\pi\Sigma}$&-&$0.049^{+0.002}_{+0.001}$\\
	$g^0_{B_0,\eta\Lambda}$&-&$-0.164^{-0.010}_{+0.014}$\\
	$g^0_{B_0,K\Xi}$&-&$-0.226^{-0.005}_{+0.079}$\\
	$m_B^0$ (MeV)&-&$1750^{+39}_{-29}$\\
	Pole 1 (MeV)&$1336-87\,i$&$1324^{-4}_{-4}-67^{+8}_{+10}\,i$\\
	Pole 2 (MeV)&$1430-26\,i$&$1428^{+4}_{+2}-24^{+4}_{+0}\,i$\\
	Pole 3 (MeV)&$1676-17\,i$&$1674^{+2}_{-4}-11^{-4}_{+3}\,i$\\
	\hline
	\hline
	\bottomrule[0.3pt]
\end{tabular}
\end{table}

Using these fits, we can obtain the $\Lambda(1670)$ pole in both scenarios. As shown in Table~\ref{coupling}, in the first scenario the pole position is located at $1676-17\,i$ MeV. This is not far from that in the second scenario, 
namely $1674 - 11\,i$ MeV.
Our results are consistent with those of other groups \cite{Oset:2001cn,GarciaRecio:2002td,Oller:2006hx,Guo:2012vv,Zhang:2013sva,Sarantsev:2019xxm,Kamano:2015hxa,Garcia-Recio:2002yxy}. The well-known two-pole structure of the $\Lambda(1405)$ is also reproduced.

The close agreement between the two scenarios for the fitted cross sections, as well as the pole positions, indicate that the present experimental data are not able to distinguish between these two very different physical pictures for the structure of the $\Lambda(1670)$. Therefore, we bring these results to the finite volume of lattice QCD and confront their predictions with lattice QCD simulation results.

\subsection{Finite volume spectrum and structure}
As discussed in the previous subsection, the scenarios with or without a bare basis state give very similar fits to contemporary experimental cross section data. That is, the present experimental data are not able to distinguish the internal structure. We shall see that the lattice QCD simulation results provide more information about this question.

By studying the finite-volume Hamiltonian with the fitted parameters given in Table~\ref{coupling}, we can obtain the corresponding lattice energy eigenvalues and eigenvectors. The lattice QCD results are provided at different pion masses, and thus we need the hadron mass dependence on the pion mass as input. For the masses of $m_{K}(m_{\pi}^2)$, $m_{N}(m_{\pi}^2)$, $m_{\Lambda}(m_{\pi}^2)$, $m_{\Sigma}(m_{\pi}^2)$ and $m_{\Xi}(m_{\pi}^2)$, we use a smooth interpolation of the corresponding lattice QCD results. The mass of the $\eta$ meson is~\cite{Wu:2018mgz} 
\begin{eqnarray}
m_\eta(m_\pi^2)=\sqrt{m_\eta^2|_{phy}+\frac{1}{3}(m_\pi^2-m_\pi^2|_{phy})}.
\end{eqnarray}
\begin{figure}[tbp]
\center
\includegraphics[width=3.4in]{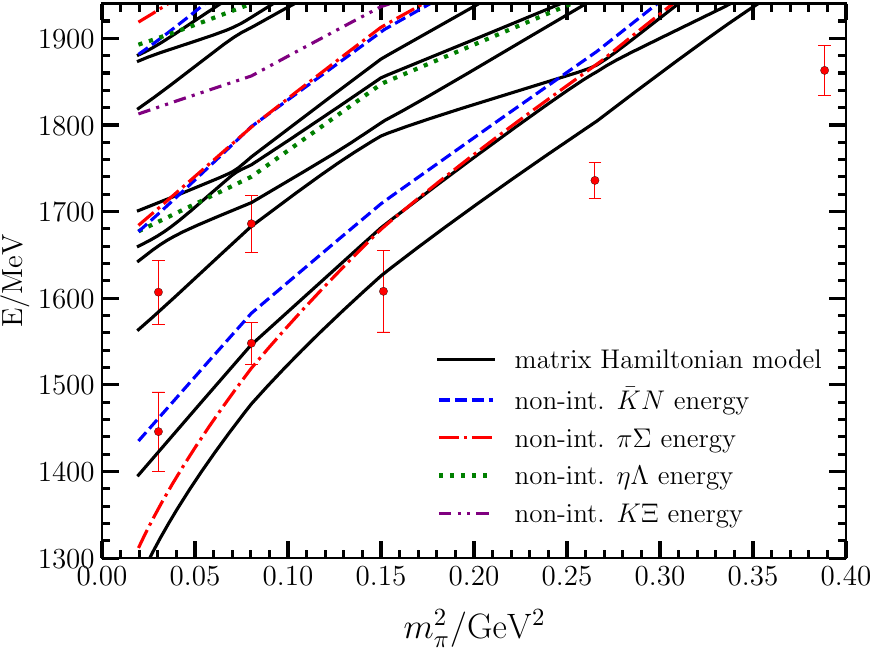}\\
(a) Without a bare $\Lambda$\\
\vspace{2em}
\includegraphics[width=3.4in]{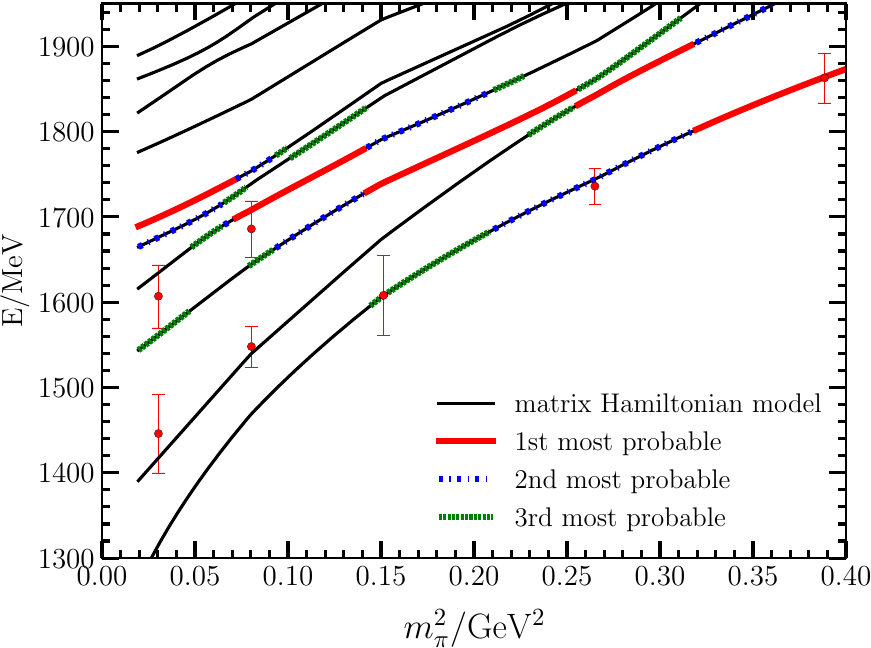}\\
(b) With a bare $\Lambda$\\
\caption{The pion mass dependence of the eigenstates obtained using the finite-volume Hamiltonian. In the upper plot, the broken lines denote noninteracting meson-baryon energies, while the solid lines denote the eigenenergies obtained from the finite-volume Hamiltonian matrix. In the lower plot, energy eigenstates based on the inclusion of a bare quark-model-like basis state are illustrated. The thick (red), dot-dashed (blue), and dotted (green) lines label the states composed with a significant contribution from the bare quark-model-like basis state, with red illustrating the largest bare state component. The negligible component of the bare basis state in the first state of the spectrum at light quark masses explains its absence in the lattice QCD spectrum excited with local three-quark operators. The lattice results are taken from the CSSM group \cite{Menadue:2011pd,Hall:2014uca} in $2+1$ flavor QCD \cite{PACS-CS:2008bkb}.}
\label{withoutbaresp}
\end{figure}
\begin{figure*}[htb]
\center
\begin{minipage}{0.48\linewidth}
	\includegraphics[width=3.9cm]{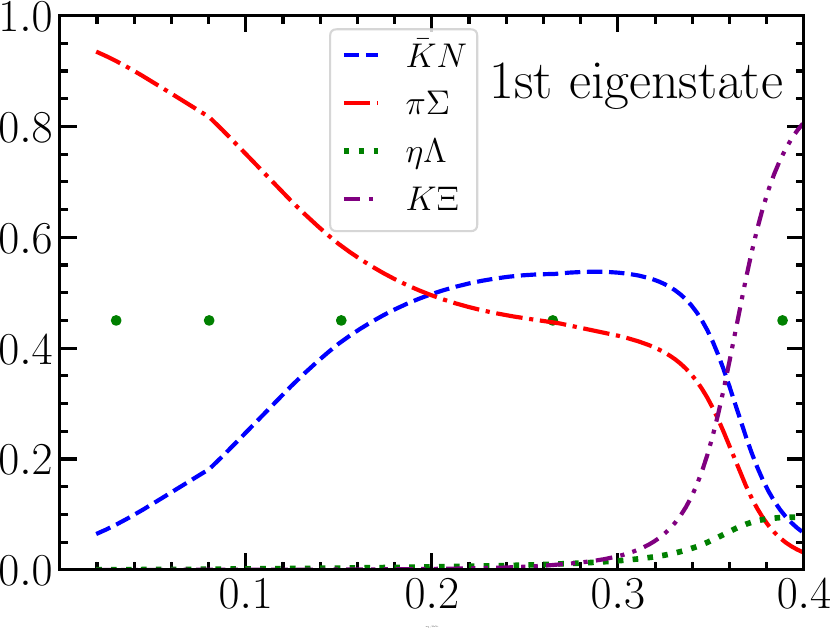}
	\includegraphics[width=3.9cm]{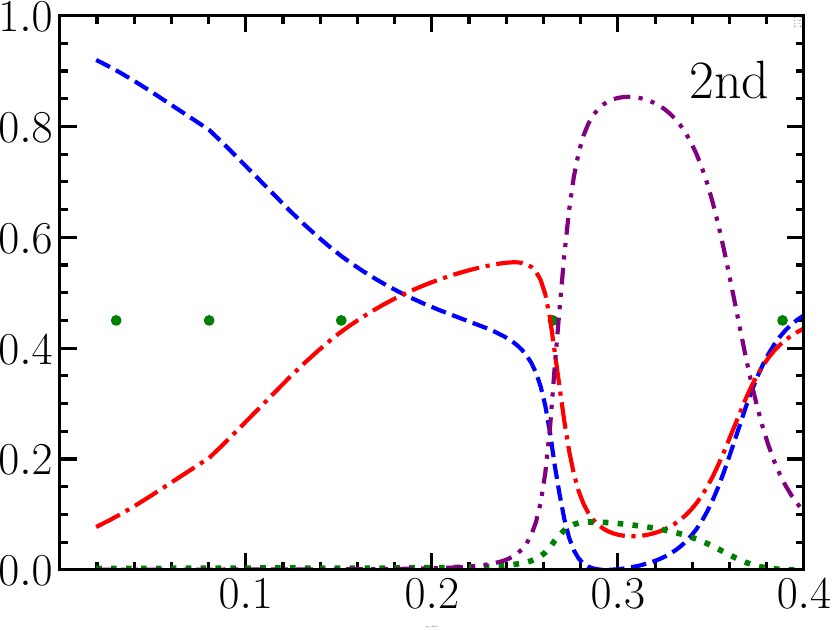}
	\includegraphics[width=3.9cm]{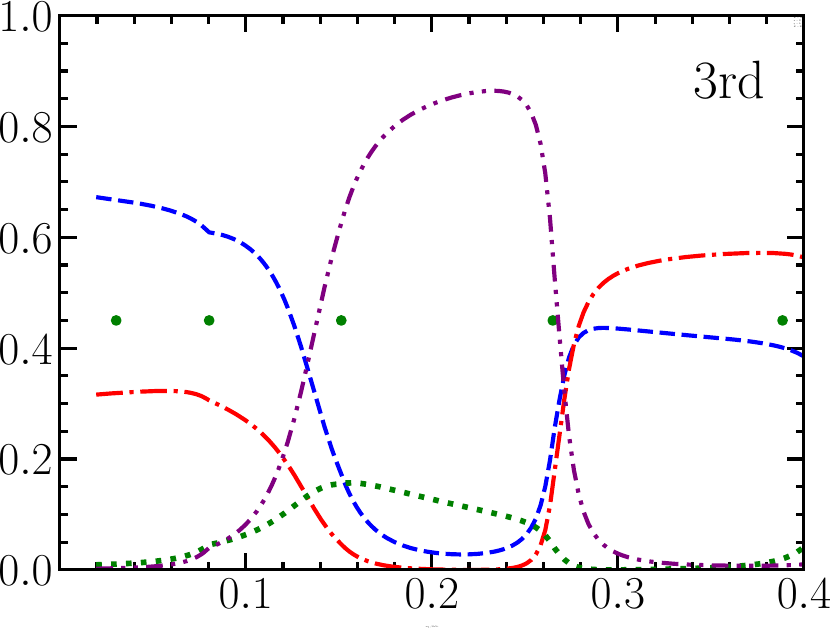}
	\includegraphics[width=3.9cm]{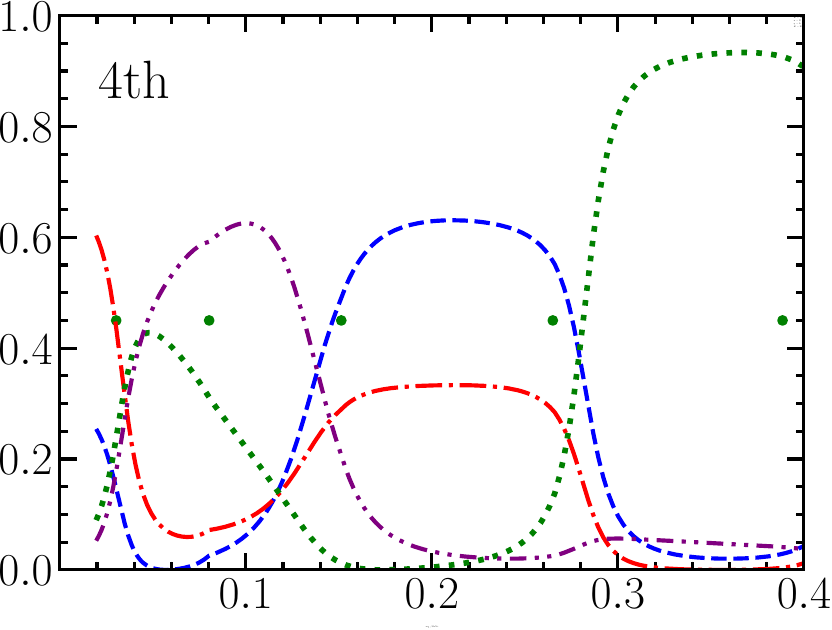}
	\includegraphics[width=3.9cm]{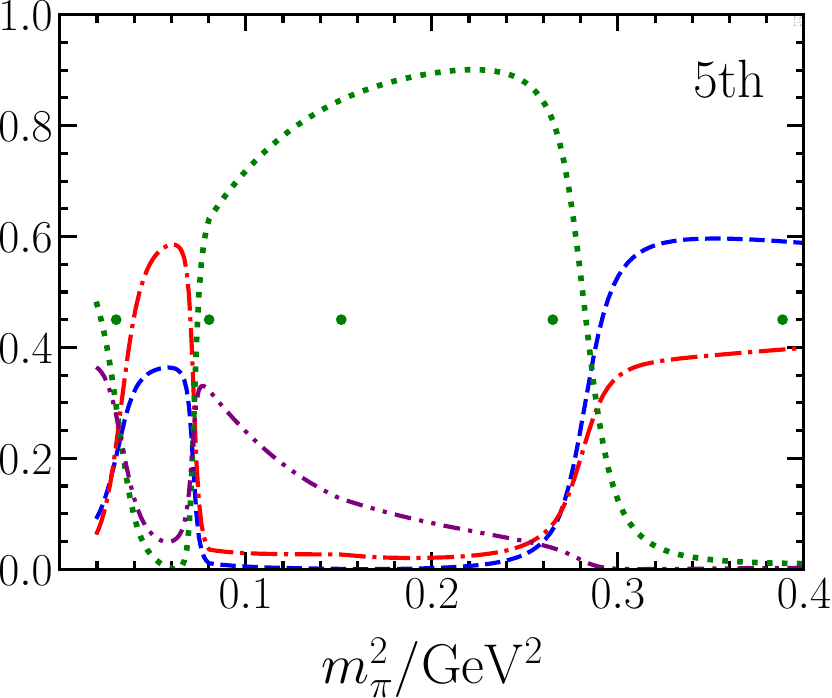}
	\includegraphics[width=3.9cm]{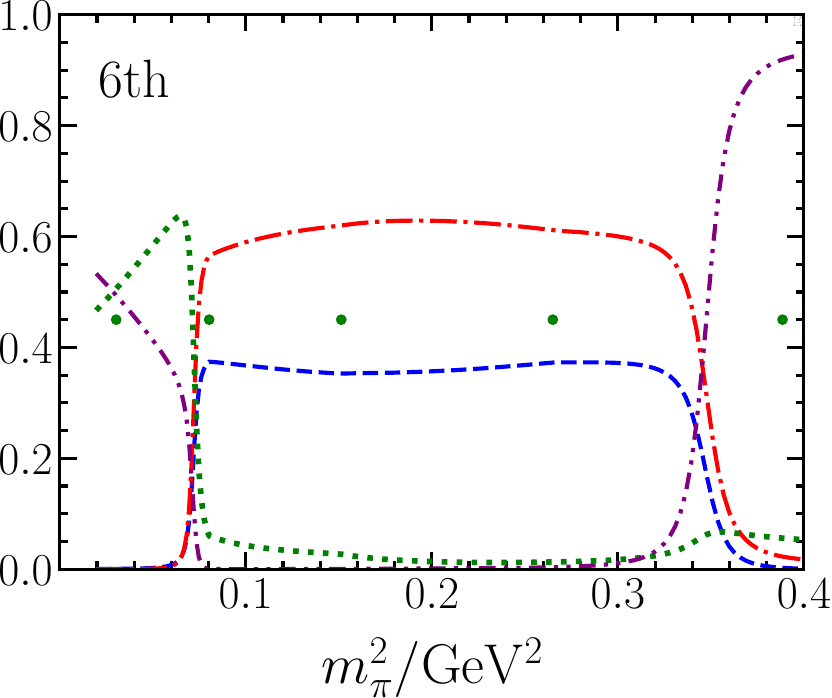}\\
	(a) Without a bare $\Lambda$\\
\end{minipage}\qquad
\begin{minipage}{0.48\linewidth}
	\includegraphics[width=3.9cm]{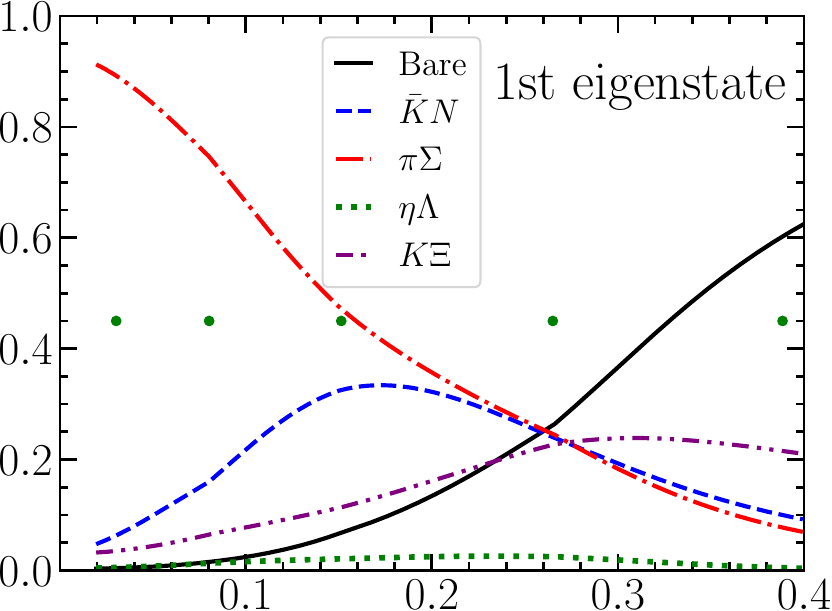}
	\includegraphics[width=3.9cm]{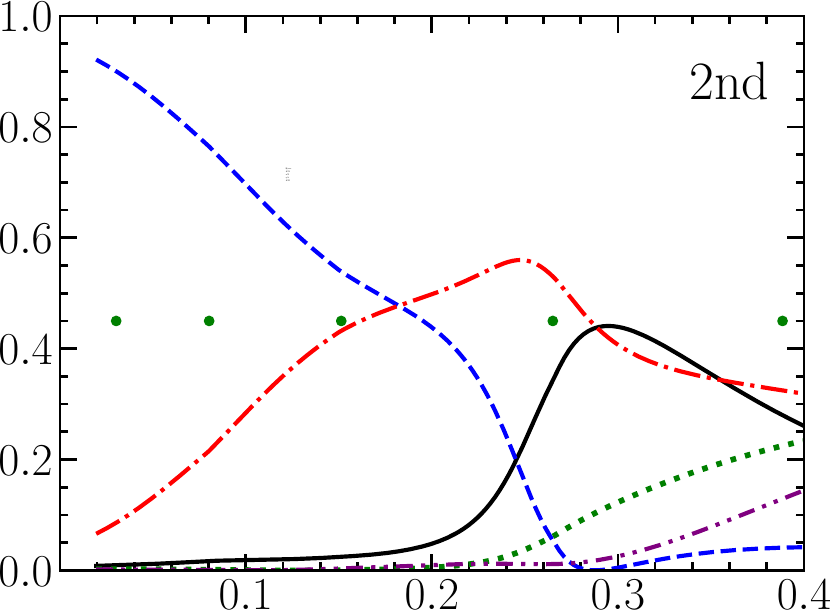}
	\includegraphics[width=3.9cm]{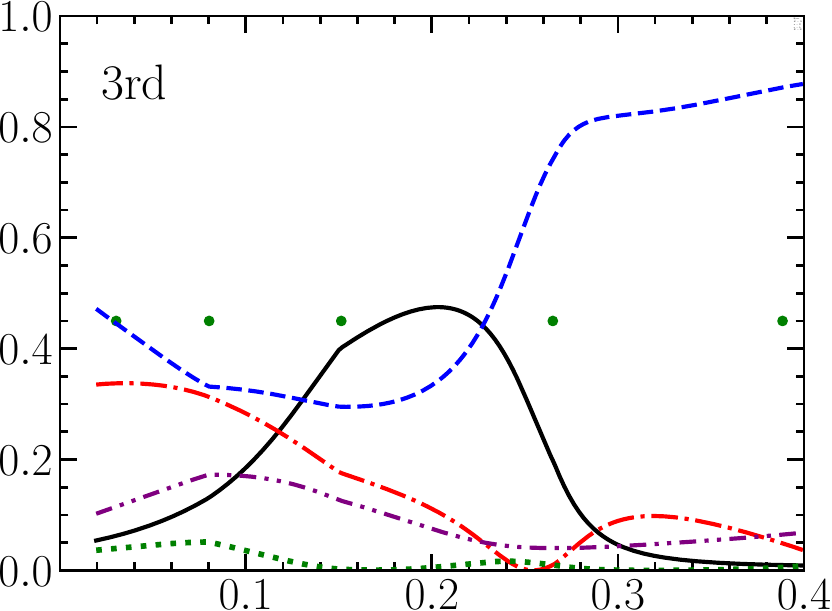}
	\includegraphics[width=3.9cm]{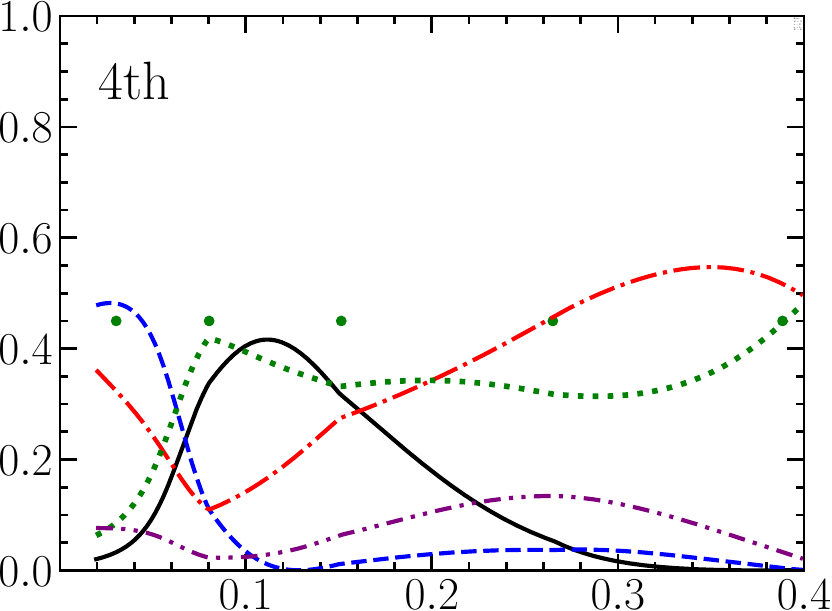}
	\includegraphics[width=3.9cm]{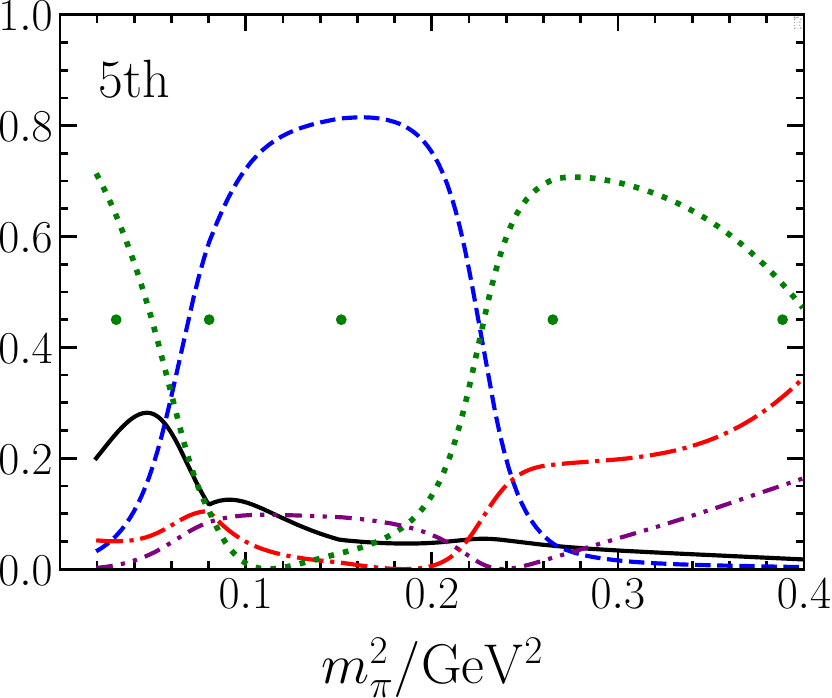}
	\includegraphics[width=3.9cm]{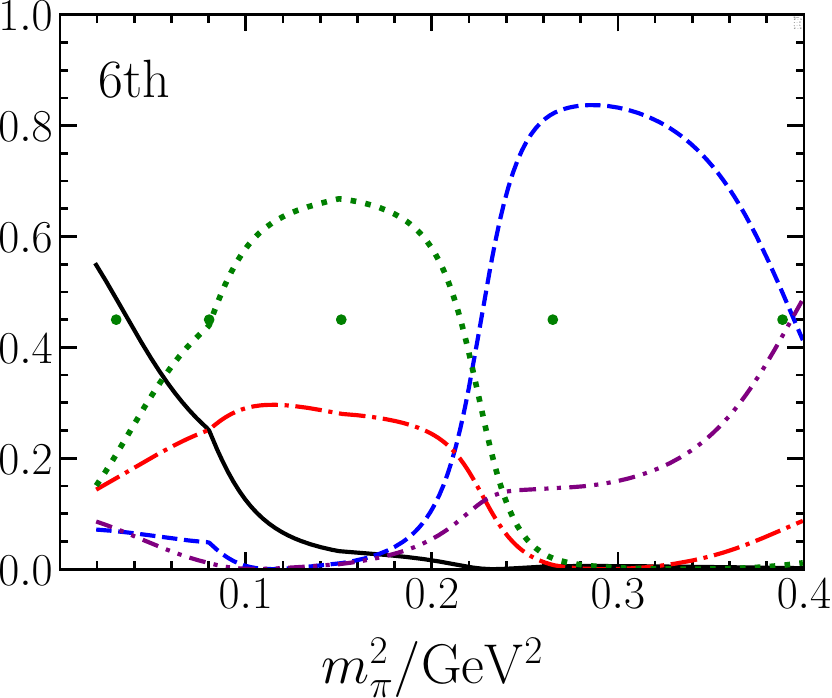}\\
	(b) With a bare $\Lambda$\\
\end{minipage}
\caption{The pion mass dependence of the Hamiltonian matrix eigenvector components for the first six states, under the assumption of no bare-baryon (left two columns) and including a bare-baryon basis (right two columns). The five circles in each diagram represent the five quark masses considered by the CSSM group \cite{Menadue:2011pd,Hall:2014uca} in $2+1$ flavor QCD \cite{PACS-CS:2008bkb}.}
\label{vectors}
\end{figure*}

We plot the pion mass dependence of the eigenstates in the finite volume Hamiltonian in Fig.~\ref{withoutbaresp} (a) for the case where the $\Lambda(1670)$ is a state without a bare baryon component. The lattice results in $\sim$3 fm box, shown as red dots with error bar, are taken from the CSSM group~\cite{Menadue:2011pd,Hall:2014uca} in $2+1$ flavor QCD~\cite{PACS-CS:2008bkb}. From Fig.~\ref{withoutbaresp} (a), we find that the results are consistent with the lattice QCD data at small pion masses, but display significant differences for the two heaviest quark masses considered.

We show the corresponding eigenvectors in Fig.~\ref{vectors} (a) for this first scenario. Near the physical pion mass, the first and second eigenstates are mainly $\pi\Sigma$ and $\bar K N$ states, respectively. The second and third eigenstates are predominantly mixtures of these two channels with $\bar K N$ continuing to dominate the third state. The fifth and sixth states are dominated by the $K\Xi$ and $\eta\Lambda$ mixture. The fifth and sixth eigenenergies in the $\sim$3 fm box are close to the position of the $\Lambda(1670)$ at the physical pion mass, as we see in Fig.~\ref{withoutbaresp} (a). With the fourth, fifth and sixth states residing in the region of the $\Lambda(1670)$ resonance, all four of the two-particle channels considered can play an important role in governing the structure of this resonance.

However, at large pion masses, these results without a bare baryon are inconsistent with 
the lattice QCD data. This was also reported in our earlier work, which focused on the  $\Lambda(1405)$~\cite{Liu:2016wxq}. From Fig.~\ref{withoutbaresp} (a), the lattice simulation at the largest pion mass is considerably lower than the first Hamiltonian eigenstate. 
This greatly reduces the probability that the odd-parity $\Lambda$ spectrum can be described by a model without the bare baryon.

To study the case with a bare quark-model-like baryon, we need to know the variation of the bare mass, $m_B^0$, as the pion mass increases. Within the quark model its mass is expected to increase linearly with the light quark mass as $m_\pi^2$ increases and hence we take
\begin{eqnarray}
m_B^0(m_{\pi}^2)=m_B^0|_{phy}+\alpha_B^0(m_{\pi}^2-m_{\pi}^2|_{phy}).
\end{eqnarray}
For the $N^*(1535)$, $\alpha_N^0=0.944$ GeV$^{-1}$ was obtained in Ref.~\cite{Abell:2023nex}.  For the $\Lambda$, where the strange quark mass is held fixed, it is natural to take 2/3 of this, such that $\alpha_B^0=0.629$ GeV$^{-1}$.

In Fig.~\ref{withoutbaresp} (b) we present the $\Lambda$ spectrum with a bare baryon basis state. Our results clearly reproduce the lattice QCD simulations well at all pion masses. The content of the corresponding eigenstates is shown in Fig.~\ref{vectors} (b). Some of this information has been brought to Fig.~\ref{withoutbaresp} (b), where colour and texture have been added to the solid lines indicating the eigenstate energies. This additional information illustrates the energy eigenstates where the bare baryon state makes a substantial contribution to the composition of the state in HEFT. The largest bare basis-state contribution is illustrated in solid red, the second largest in dot-dash blue and the third largest contribution in dotted green. 

In the first eigenstate, the main component is $\pi\Sigma$ at small pion masses, while the contributions of $\pi\Sigma$ and $\bar{K}N$ channels become comparable and then the bare baryon dominates as the pion mass becomes larger. The second eigenstate is mainly composed of $\bar{K}N$, while the third and fourth are dominated by the $\bar{K}N$ and $\pi\Sigma$ channels at small pion masses with $\bar KN$ continuing to dominate for both states. At the physical pion mass, the fifth state is dominated by $\eta \Lambda$ with a significant bare baryon component.  
The sixth state is dominated by the quark-model-like basis state. Remarkably, all four of
the two-particle channels provide the balance of basis-state contributions at the physical
point.

With the bare basis state contributing in the $\Lambda(1670)$ region, it is now clear the
CSSM collaboration was able to excite the $\bar K N$ state with a local three-quark
operator due to its localised structure. While the strange magnetic form factor shows the
contribution of a vacuum quark-antiquark pair to create a 5-quark $\bar K N$ state
\cite{Hall:2014uca}, the electric form factors describe a localised state. Figure 3 of Ref.~\cite{Menadue:2013xqa} shows the strange quark distribution is largely unchanged
between the ground-state positive-parity $\Lambda$ and its first excitation in the $\Lambda(1405)$ region.  Similarly, the light-quark distribution
grows only slightly from the ground state to the $\bar K N$ state in the $\Lambda(1405)$ resonance regime.

In summary, the lattice QCD results favor the scenario in which the $\Lambda(1670)$
contains a bare-baryon component. As the fifth and sixth states in this description contain
very significant quark-model-like basis state contributions and because the fourth, fifth and
sixth states sit in the $\Lambda(1670)$ resonance regime, as illustrated in Fig.~\ref{withoutbaresp}(b), one
can conclude that the $\Lambda(1670)$ has a quark-model-like core dressed by all four of
the meson-baryon channels considered.

\subsection{Uncertainty analysis}
\label{sec:uncertaintyAnalysis}

\begin{figure}[tbp]
\center
\includegraphics[width=3.4in]{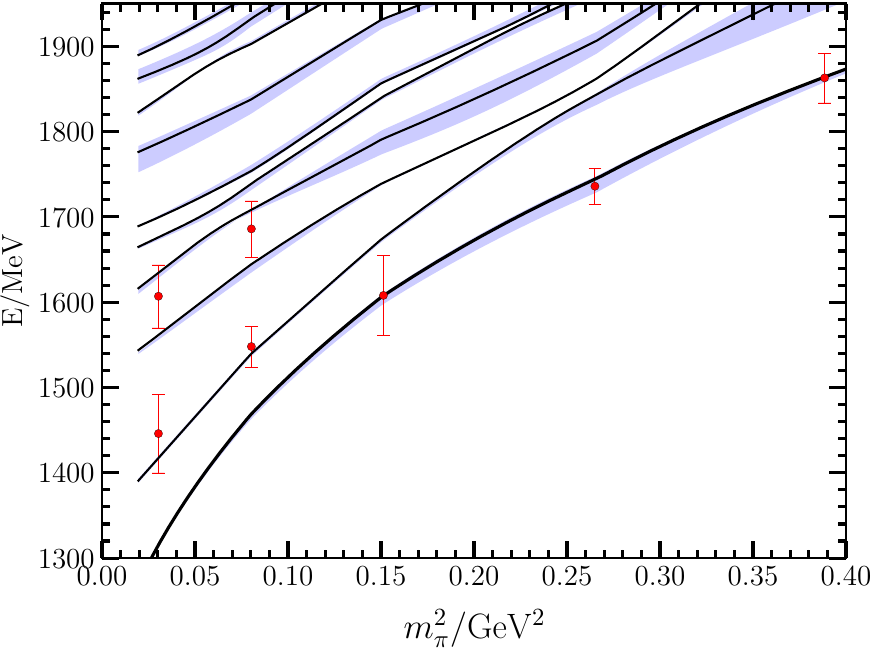}
\caption{Error estimation for the spectrum of odd-parity strange spin-1/2 baryons in HEFT for the CSSM lattice volume. The black solid lines represent the values with the optimal Hamiltonian parameters while the blue shaded regions illustrate the uncertainty in the HEFT results obtained through the allowed variation of Hamiltonian parameters as described in Sec.~\ref{sec:uncertaintyAnalysis}.}
\label{figError}
\end{figure}

To obtain an error estimate on the link between experiment and the
finite-volume spectra driven by the embedded L\"uscher relation, we
draw on the regularisation parameter, $\Lambda$, to move in the
Hamiltonian parameter space and explore alternative mediations between
experiment and theory.

As noted in the previous section, the constraints of experiment and
lattice QCD are effective in constraining the Hamiltonian parameters,
allowing only a small range of variation in $\Lambda$. If one forces
$\Lambda$ outside of the range allowed by the experimental data, the
fit to the data is spoiled and thus the associated finite-volume
energy spectrum becomes incorrect. In a similar manner, the correct
description of lattice QCD results places constraints on the variation
of parameters.

We commence by changing $\Lambda$ by 50 MeV from our initial value of
1.0 GeV and refitting the parameters to describe experiment. This
small variation is repeated, monitoring the $\chi^2$ per degree of
freedom to ensure the experimental data continues to be described in an
accurate manner.  The finite volume spectrum is then calculated.  We
compare the results with the CSSM lattice QCD results to ensure a
valid description of the lattice QCD constraint.

The variation of the $\chi^2$/dof for the cross section fits is subtle
over the range $0.90 \le \Lambda \le 1.10$ GeV but jumps significantly
for the values $\Lambda = 0.85$ and 1.15 GeV.  On this basis alone,
the fits for $\Lambda$ $<$ 0.85 GeV and $\Lambda > 1.15$ GeV are
excluded. However, considering the lattice QCD constraint further
excludes $\Lambda > 1.15$ GeV.  Over the rage $0.90 \le \Lambda \le
1.10$ GeV the three pole positions do not change by more than 10 MeV.

The best description of the lattice QCD results is provided by
$\Lambda = 1.00$ GeV and we refer to this for our central values.  To
produce uncertainties in the finite-volume results, we refer to the
predictions for $\Lambda = 0.90$ and 1.10 GeV and use these results to
shade error bars in Figs.~\ref{figError} and \ref{BaScCom}.  Uncertainties in the fit parameters of Table \ref{coupling} also follow from this range of allowed $\Lambda$ variation.

\subsection{Comparison with the latest lattice QCD simulation}
One can clearly see that some eigenstates predicted by the HEFT are absent in the lattice QCD simulations of the CSSM group from Fig.~\ref{withoutbaresp}. More than ten years have passed since that odd-parity $\Lambda$ spectrum was obtained~\cite{Menadue:2011pd} and lattice QCD techniques have improved. With the parameters of the Hamiltonian constrained by experimental data and the results from one lattice QCD collaboration, we can now proceed to make predictions for the finite-volume spectra observed in other lattice QCD calculations, both at different volumes and at different quark masses. Very recently, the BaSc collaboration presented their coupled-channel simulations with both single baryon and meson-baryon interpolating operators in a larger box at $m_\pi\approx$200 MeV \cite{BaryonScatteringBaSc:2023zvt,BaryonScatteringBaSc:2023ori}. We now compare our HEFT predictions with this latest lattice QCD simulation.  
\begin{figure}[tbp]
\center
\includegraphics[width=2.5in]{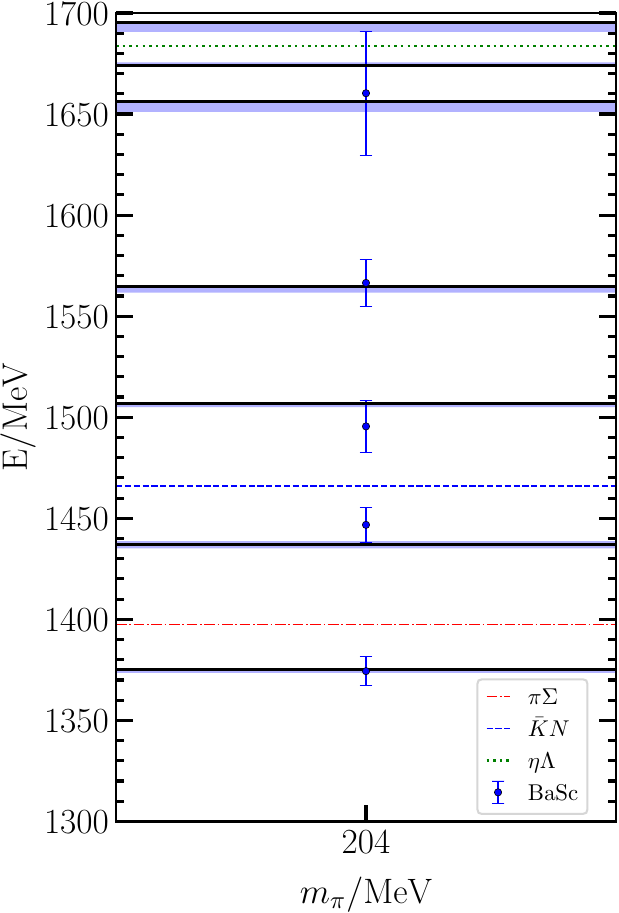}
\caption{The energy eigenvalues calculated in HEFT in the scenario including a quark-model-like single-particle basis state (solid black lines) are compared with lattice QCD calculations from the BaSc collaboration \cite{BaryonScatteringBaSc:2023zvt,BaryonScatteringBaSc:2023ori} in the $G_{1u}(0)$ irreducible representation (data points) on a $L= 4.05$ fm lattice. Dashed lines indicate the meson-baryon thresholds.
	The blue shaded regions illustrate the uncertainty in the HEFT predictions obtained through the allowed variation of Hamiltonian parameters as described in Sec.~\ref{sec:uncertaintyAnalysis}.
}
\label{BaScCom}
\end{figure}

We use the corresponding hadron masses at $m_\pi\sim 200$ MeV as reported in the lattice QCD 
simulations~\cite{BaryonScatteringBaSc:2023zvt,BaryonScatteringBaSc:2023ori} and give our HEFT results, including  the bare baryon, in Fig.~\ref{BaScCom}. One can see that the HEFT results describe the BaSc simulations very well. The lowest data point has a very small error bar but sits exactly on our lowest-lying odd-parity state. All of the HEFT energy eigenstates are far from the noninteracting meson-baryon thresholds but still coincide with the lattice results. We stress that no parameters have been adjusted in making the finite volume predictions for the BaSc lattice results.

In our approach, at $m_\pi = 204$ MeV, the first and second states observed in this $\sim4$ fm box are mainly $\pi\Sigma$ and $\bar K N$ states, respectively. The third and fourth ones are the $\pi\Sigma$-$\bar K N$ mixtures. The fifth eigenstate contains $\bar K N$ and $\pi\Sigma$ with some bare baryon. The sixth eigenstate is dominated by $\eta \Lambda$ mixed with a small component of the bare baryon. Noting that the fifth and sixth energy eigenstates are in the $\Lambda(1670)$ resonance regime, one can once again conclude that the $\Lambda(1670)$ is composed of a single-particle quark-model-like core dressed by the isoscalar meson-baryon channels considered.

\section{Summary}
\label{sec4}
In this work we have studied two different scenarios for the internal structure of the $\Lambda(1670)$. One scenario assumes that the $\Lambda(1670)$ is dynamically generated through rescattering between the $\bar{K}N$, $\pi\Sigma$, $\eta\Lambda$ and $K\Xi$ channels with $I=0$. The other assumes that the $\Lambda(1670)$ is a bare quark-model-like basis state mixing with these $I=0$ interacting channels. We fit the experimental cross section data for the $K^-p\rightarrow K^-p$, $K^-p\rightarrow \bar{K}^0n$, $K^-p\rightarrow \pi^0\Lambda^0$, $K^-p\rightarrow \pi^-\Sigma^+$, $K^-p\rightarrow \pi^0\Sigma^0$, $K^-p\rightarrow \pi^+\Sigma^-$, and $K^-p\rightarrow \eta\Lambda^0$ reactions, with the laboratory momentum of the anti-kaon in the range 0-800 MeV/c, including the recently measured threshold cross sections which have small error 
bars~\cite{Piscicchia:2022wmd}. Our fits are consistent with the cross section data if we neglect the effect of the $D$-wave $\Lambda(1520)$ resonance. In addition, we have checked the two-pole structure of the $\Lambda(1405)$ and obtained the pole position of the $\Lambda(1670)$. All of these results are consistent with those of other groups.

It is clear from the quality of the fits to the cross section data under both scenarios that one cannot distinguish between them using scattering data alone. This serves as motivation to introduce HEFT to further explore the structure of the $\Lambda(1670)$ in the finite volume of lattice QCD. The scenario without a bare baryon is inconsistent with the lattice QCD data at large pion masses. Without adjusting any other parameter, the scenario including a bare-baryon basis state yields an excellent description of the lattice QCD results over the full range of light quark mass. Our HEFT results also  agree very well with 
the latest BaSc lattice QCD simulation results at $m_\pi = 204$ MeV.
Not only are the predicted energy levels very close to those reported by BaSc, but all five of the lowest eigenstates predicted in HEFT were observed in the lattice calculations.

Based on the present HEFT analysis, the lattice QCD calculations provide invaluable information about the structure of the $\Lambda(1670)$. It definitely contains a considerable single-particle quark-model-like basis state component, which mixes with the meson-baryon channels. While our calculations could be extended by considering the $\Lambda(1800)$ resonance as well as $\bar K^* N$ and  $\pi\Sigma^*$ channels, the main conclusion in this work is not expected to be sensitive to extensions well beyond the $\Lambda(1670)$ resonance regime.

\section*{Acknowledgments}
This work was supported by the National Natural Science Foundation of China under Grants No. 12175091, No. 12335001, No. 12247101, No. 12305090 and No. 12347119, the China National Funds for Distinguished Young Scientists under Grant No. 11825503, National Key Research and Development Program of China under Contract No. 2020YFA0406400, the 111 Project under Grant No. B20063, the innovation project for young science and technology talents of Lanzhou city under Grant No. 2023-QN-107, the fundamental Research Funds for the Central Universities, and the project for top-notch innovative talents of Gansu province. D. G. is also supported by the China Postdoctoral Science Foundation under Grant No. 2023M740117. This research was undertaken with the assistance of resources from the National Computational Infrastructure (NCI), provided through the National Computational Merit Allocation Scheme, and supported by the Australian Government through Grant No.~LE190100021. This research was supported by the University of Adelaide and by the Australian Research Council through ARC Discovery Project Grants No. DP190102215 and No. DP210103706 (D.B.L) and No. DP230101791 (A.W.T).


\end{document}